\documentclass[onecolumn,secnumarabic,amssymb, nobibnotes, aps, prd]{revtex4}
\usepackage{amsmath} \usepackage{amssymb}
\usepackage{graphicx} 
\usepackage{epstopdf}
\usepackage{amsmath}
\usepackage{amsmath,amssymb,amsthm,amsfonts,mathrsfs,bm,verbatim}
\usepackage{graphicx,subfigure}

\newcommand{\bea}{\begin{eqnarray}}
\newcommand{\eea}{\end{eqnarray}}

\begin{document}

\title{Analysis and research on superradiant stability of Kerr-sen Black Hole }%

\author{Wen-Xiang Chen}
\affiliation{Department of Astronomy, School of Physics and Materials Science, GuangZhou University, Guangzhou 510006, China}
\email{wxchen4277@qq.com}

\begin{abstract}
 Kerr-Sen black holes have stretchon parameters and hidden conformal symmetries. The superradiation stability and steady-state resonance are worth further study. This is the research motivation of this paper.In that article, a new variable y is added here to expand the results of the above article. When$\sqrt{2a^2}/{r^2_+}< \omega< m\varOmega_H+q\varPhi_H$,so the Kerr-sen black hole is superradiantly stable at that time,similar to the superradiation result of the Kerr-Newman black hole.
\end{abstract}

\maketitle

\section{Introduction}
The No Hair Theorem of black holes first proposed in 1971 by Wheeler and proved by Stephen Hawking\cite{1}, Brandon Carter,  etc, in 1973. In 1970s, the development of black hole thermodynamics applies the basic laws of thermodynamics to the theory of black hole in the field of general relativity, and strongly implies the profound and basic relationship among general relativity, thermodynamics and quantum theory. 
The stability of black holes is an major topic in black hole physics. Regge and Wheeler\cite{2} have proved that the spherically symmetric Schwarzschild black hole is stable under perturbation. The great impact of superradiance makes the stability of rotating black holes more complicated. Superradiative effects occur in both classical and quantum scattering processes. When a boson wave hits a rotating black hole, chances are that rotating black holes are stable like Schwarzschild black holes, if certain conditions are satisfied\cite{1,2,3,4,5,6,7,8,9}
\begin{equation}
    \omega< m\varOmega_H+q\varPhi_H,  \varOmega_H=\frac{a}{r_+^2+a^2}
\end{equation}
where $q$ and $m $ are the charge and azimuthal quantum number of the incoming wave, $\omega$ denotes the wave frequency, $\varOmega_H$ is the angular velocity of black hole horizon and $\varPhi_H$ is the electromagnetic potential of the black hole horizon. If the frequency range of the wave lies in the superradiance condition, the wave reflected by the event horizon will be amplified, which means the wave extracts rotational energy from the rotating black hole when the incident wave is scattered. According to the black hole bomb mechanism proposed by Press and Teukolsky\cite{1,2,3,4,5,6,7,8,9}, if a mirror is placed between the event horizon and the outer space of the black hole, the amplified wave will reflect back and forth between the mirror and the black hole and grow exponentially, which leads to the super-radiative instability of the black hole.

 The phenomenon of superradiation allows the extraction of rotational and Coulomb energies from rotating or charged black holes. In 1972, Press and Teukolsky\cite{5,6,7,8,9} proposed adding a mirror to the outside of a black hole, using a superradiation mechanism to make a black hole bomb. Later, it was found that the mass term of scalar field, anti-de-Sitter (AdS) space-time boundary, Kaluza-Klein momentum, cosmological constant, infinite high potential barrier outside linear stretchon black hole, etc., can effectively act as a reflecting mirror (the function of mirror can effectively prevent scattered waves from escaping to infinity, It can be provided by man or nature itself). Under this mechanism, the superradiation wave mode will reflect back and forth between the black hole horizon and the mirror and be amplified exponentially, so as to continuously extract the rotational energy and Coulomb energy of the black hole and lead to the instability of the black hole, which is called the superradiation instability and produce the black hole bomb effect.

In Kerr black holes, the qualitative theory of superradiation of scalar field, electromagnetic field and gravitational wave was first given by Starobinsky's classic paper\cite{5,6,7,8,9}. Due to the existence of the energy layer of the rotating black hole, the negative energy flow of the scattered waves in the black hole will enter the interior of the energy layer, and the energy conservation, the energy of the black hole will be extracted by the outgoing waves, making the amplitude increase. This provides the basis for a physical explanation of the superradiation of black holes.

The superradiation stability of black holes has always been an open research topic. The classical superradiative stability and steady-state resonance problems of Kerr black hole, Kerr-Newman black hole, Kerr-Newman-anti-de Sitter black hole, Reissner-Nordstrom Black hole, Kerr-Godel black hole, and rotational linear contractor black hole, In recent years, it has been extensively and deeply studied by many authors, such as S. Hod, Ran Li, V. Cardoso, A.N. Aliev, etc., using analytical and numerical methods\cite{10,11,12,13,14}. The space-time of Kerr-Sen black holes obtained from string theory is different from that of general relativity, so it attracts great attention. However, as far as the author knows, no one has studied the superradiative stability and steady-state resonance of Kerr-Sen black hole so far. Kerr-sen black hole has the properties of rotation, charge and symmetry as Kerr-Newman black hole and other asymptotically flat space-time. Moreover, Kerr-Sen black holes have stretchon parameters and hidden conformal symmetries. The superradiation stability and steady-state resonance are worth further study. This is the research motivation of this paper.

In that article\cite{43}, a new variable y is added here to expand the results of the above article. When $\mu \ge \sqrt{2}(m\varOmega_H+q\varPhi_H)$,so the Kerr-Newman black hole is superradiantly stable at that time.

In this article, a new variable y($\mu=y\omega$) has been added to extend the results of the above article. We use the property of curve integrals. When y is greater than some limit, the effective potential of the equation has no poles, then there is no potential well beyond the event horizon,  when $\sqrt{2a^2}/{r^2_+}< \omega< m\varOmega_H+q\varPhi_H$,so\ the Kerr-sen black\ hole\ is\ superradiantly\ stable\ at\ that\ time.

\section{The system of Kerr-Newman black hole}

 The Kerr-sen solution is the strict classical four-dimensional solution in the Boyer Lindquist coordinate system $(t, r, \theta, \phi)$in which the kerr-sen space-time metric can be written (with natural unit, $G=\hbar=c=1$)\cite{11,12,13,14,15,16,17,18,19,20,21,22,23,24,36}
\begin{equation}
\begin{aligned}
d s^{2} &=-\left(1-\frac{2 M r}{\rho^{2}}\right) d t^{2}+\rho^{2}\left(\frac{d r^{2}}{\Delta}+d \theta^{2}\right) \\
&-\frac{4 M r a}{\rho^{2}} \sin ^{2} \theta d t d \phi+\left[r(r+2 b)+a^{2}+\frac{2 M r a^{2} \sin ^{2} \theta}{\rho^{2}}\right] \sin ^{2} \theta d \phi^{2}
\end{aligned}
\end{equation}
where
\begin{equation}
A_{\mu}=-\frac{Q r}{\rho^{2}}\left(d t-a \sin ^{2} \theta d \phi\right)
\end{equation},
\begin{equation}
\begin{aligned}
&\Delta \equiv r^{2}-2 M^{\prime} r+a^{2} \\
&\rho^{2}=r^{2}+2 b r+a^{2} \cos ^{2} \theta \\
&\Sigma=r^{2}+2 b r+a^{2} \\
&r_{\pm}=M^{\prime} \pm \sqrt{M^{\prime2}-a^{2}} \\
&M^{\prime}=M-b=M-\frac{Q^{2}}{2 M}
\end{aligned}
\end{equation}
$r_{\ PM}$is the inner and outer horizon of the black hole, and $b$ is the parameter related to the extender, which satisfy:
\begin{equation}
D=-\frac{1}{2} \ln \frac{r^{2}+2 b r+a^{2} \cos ^{2} \theta}{r^{2}+a^{2} \cos ^{2} \theta}
\end{equation}

We study the physical and mathematical properties of linearized massive scalar field configurations (scalar clouds) nontrivially coupled to the electromagnetic field of Kerr-Sen black holes. The space-time line element of a spherically symmetric Kerr-Sen black hole can be expressed as

\begin{equation}
d s^{2}=-g(r)dt^2+{\frac{1}{g(r)}}dr^2+r^2(d\theta^2+\sin^2\theta
d\phi^2) ,
\end{equation}
where
\begin{equation}\label{Eq2}
g(r)=1-{\frac{2M^{\prime}}{r}}+{\frac{a^2}{r^2}}\  .
\end{equation}

 The following covariant Klein-Gordon equation
\begin{equation}
( \nabla ^{\nu}-iqA^{\nu})( \nabla _{\nu}-iqA_{\nu}) \Phi =\mu ^2\Phi,
\end{equation}
where $\nabla ^{\nu}$ represents the covariant derivative under the Kerr-Newman background. We adopt the method of separation of variables to solve the above equation, and it is decomposed as
\begin{equation}
\Phi ( t,r,\theta ,\phi) =\sum_{lm}R_{lm}( r ) S_{lm}( \theta) e^{im\phi}e^{-i\omega t}.
\end{equation}
where $R_{lm}$ are the equations which satisfy the radial equation of motion. The angular function $S_{lm}$ denote the scalar spheroidal harmonics which satisfy the angular part of the equation of motion. $l(=0,1,2,...)$ and $m$ are integers, $-l\leq m\leq l$ and $\omega$ denote the angular frequency of the scalar perturbation. 

The angular part of the equation of motion is an ordinary differential equation and it can be expressed as follows, 
\begin{equation}
\frac{1}{\sin \theta} \frac{d}{\theta}\left(\sin \theta \frac{d S_{l m}}{d \theta}\right)+\left[K_{l m}+a^{2}\left(\mu^{2}-\omega^{2}\right)-a^{2}\left(\mu^{2}-\omega^{2}\right) \cos ^{2} \theta-\frac{m^{2}}{\sin ^{2} \theta}\right] S_{l m}=0
\end{equation}
where $K_{lm}$ represent angular eigenvalues. The standard spheroidal differential equation above has been studied for a long time  and of great significance in a great deal of physical problems. The spheroidal functions $S_{lm}$ are known as prolate (oblate) for $( \mu ^2-\omega ^2) a^2>0( <0)$, and only the prolate case is concerned in this article. 
We select the lower bound for this separation constant as follow,\cite{25,26,27,28,29,30,31,32,33,34,35}
\begin{equation}
K_{l m}+a^{2}\left(\mu^{2}-\omega^{2}\right)=l(l+1)+\sum_{k=1}^{\infty} c_{k} a^{2 k}\left(\mu^{2}-\omega^{2}\right)^{k}
\end{equation}

The radial part of the Klein-Gordon equation contented by $R_{lm}$ is written as
\begin{equation}
\frac{d}{d r}\left(\Delta \frac{d R_{l m}}{d r}\right)+\left(\frac{H^{2}}{\Delta}+2 m a \omega-\mu^{2} \Sigma-K_{l m}\right) R_{l m}=0
\end{equation}
where
\begin{equation}
H=\omega\left(r^{2}+2 b r+a^{2}\right)-a m-q Q r
\end{equation}.The radial equation is converted to
\begin{equation}
\frac{d^{2} U}{d y^{2}}+\frac{\Delta}{\Sigma^{3 / 2}}\left[\frac{1}{\Sigma^{1 / 2}}\left(\frac{H^{2}}{\Delta}+2 m a \omega-\mu^{2} \Sigma-K\right)-\frac{1}{2} \frac{d}{d r}\left(\frac{\Delta}{\Sigma^{3 / 2}} \frac{d \Sigma}{d r}\right)\right] U=0
\end{equation}

We get the following radial wave equation(another\ radial\ function\ $ \psi =\text{rR}$)

\begin{equation}
\frac{{{\text{d}}^{2}}\psi }{\text{dr}_{*}^{\text{2}}}+V\psi =0,
\end{equation},
V can change to \cite{17}
\begin{equation}
V(r)=\left(1-\frac{2M^{\prime}}{r}+\frac{({a}^{2})}{r^{2}}\right)\left[\mu^{2}+\frac{l(l+1)}{r^{2}}+\frac{2 M^{\prime}}{r^{3}}-\frac{2 ({a}^{2})}{r^{4}}-\frac{\alpha ({a}^{2})}{r^{4}}\right]
\end{equation}

It is necessary to consider asymptotic solutions of radial equations near the event horizon and at infinity of space under suitable boundary conditions to study the superradiation modes of Kerr-Sen black holes with charged massive scalar perturbations. In this paper, the boundary conditions of radial functions are analyzed by using turtle coordinate method.
The tortoise coordinate $r_*$ is defined by the following equation
\begin{equation}
\frac{dr_*}{dr}=\frac{r^2+a^2}{\varDelta}.
\end{equation}
The two boundary conditions we are concerned with are pure incoming waves near the outer horizon and exponentially decaying waves at infinity in space. Therefore, the asymptotic solution of the radial wave function under the above boundary is selected as follows
\begin{equation}
R(y)= \begin{cases}A e^{i \sigma r_*}+B e^{-i \sigma r_*}, & r \rightarrow r_{+}(r_* \rightarrow-\infty) \\ C e^{i \sqrt{\omega^{2}-\mu^{2}} y}+D e^{-i \sqrt{\omega^{2}-\mu^{2}}r_*}, & r \rightarrow+\infty(r_* \rightarrow+\infty)\end{cases}
\end{equation}
We can easily see that getting decaying modes at spatial infinity requires following bound state condition
\begin{equation}\label{bsc}
\omega ^2<\mu^2.
\end{equation}
The critical frequency $\omega_c$ is defined as
\begin{equation}
\omega _c=m\varOmega _H+q\varPhi _H,
\end{equation}
where
\begin{equation}
\begin{aligned}
&\sigma=\omega-\omega_{c} \\
&\omega_{c} \equiv m \Omega_{\mathrm{H}}+q \Phi_{\mathrm{H}} \\
&\Omega_{\mathrm{H}}=\frac{a}{r_{+}^{2}+2 b r_{+}+a^{2}} \\
&\Phi_{\mathrm{H}}=\frac{Q r_{+}}{r_{+}^{2}+2 b r_{+}+a^{2}}
\end{aligned}
\end{equation}
$\Omega_{H}$and $\Phi_{H}$are the angular velocities and electromagnetic potentials at the outer horizon of the black hole, respectively.

\section{The radial equation of motion and effective potential}

A new radial wave function is defined as\cite{11,30,31,32,33,34,35}
\begin{equation}
\psi _{lm}\equiv \varDelta^{\frac{1}{2}}R_{lm}.
\end{equation}
in order to substitute the radial equation of motion for a Schrodinger-like equation
\begin{equation}
\frac{d^2\Psi _{lm}}{dr^2}+( \omega ^2-V1) \Psi _{lm}=0,
\end{equation}

Taking the superradiant condition, i.e. $\omega<\omega_c$, and bound state condition into consideration,  the  Kerr-sen black hole and charged massive scalar perturbation system are superradiantly stable when the trapping potential well outside the outer horizon of the  Kerr-sen black hole does not exist. As a result, the shape of the effective potential $V$ is analyzed next in order to inquiry into the nonexistence of a trapping well.

The asymptotic behaviors of the effective potential $V$ around the inner and outer horizons and at spatial infinity can be expressed as
\begin{equation}
V1( r\rightarrow +\infty )
\rightarrow \mu ^2-\frac{2( 2M^{\prime}\omega ^2-M^{\prime}\mu ^2)}{r}+{\cal O}( \frac{1}{r^2}) ,
\end{equation}
\bea
V1( r\rightarrow r_+ ) \rightarrow -\infty,~~
V1( r\rightarrow r_-) \rightarrow -\infty.
\eea

If a Kerr black hole satisfy the condition of $\mu=y\omega$, it will be superradiantly stable when $\mu<\sqrt{2}m\Omega_H$. In this article, we introduce the above condition into  Kerr-sen black holes. Therefore, the formula of the asymptotic behaviors is written as

\begin{equation}
V1( r\rightarrow +\infty )
\rightarrow y^2\omega^2-\frac{2[M^{\prime}(2-y^2)\omega ^2] }{r}+{\cal O}( \frac{1}{r^2}) ,
\end{equation}
\bea
V1( r\rightarrow r_+ ) \rightarrow -\infty,~~
V1( r\rightarrow r_-) \rightarrow -\infty.
\eea
It is concluded from the equations above that the effective potential approximates a constant at infinity in space, and the extreme between its inner and outer horizons cannot be less than one.The asymptotic behaviour of the derivative of the effective potential $V$ at spatial infinity can be expressed as 
\begin{equation}
 V1'( r\rightarrow +\infty )
 \rightarrow \frac{2[ M^{\prime}(2-y^2)\omega ^2 ]}{r^2}+{\cal O}( \frac{1}{r^3}) ,
\end{equation}
The derivative of the effective potential has to be negative in order to satisfy the no trapping well condition,
\begin{equation}
2 M^{\prime}(2-y^2)\omega^2<0.
\end{equation}

\section{The limit $y$ of the incident particle under the superradiance of Kerr-sen black holes}
Gravity is traditionally used to simulate the phenomenon of superradiation, which is expected to realize the space-time rotation of toy models. A widely considered configuration is based on whirlpools, such as draining bathtubs; Recently, the first experimental evidence of superradiative scattering was obtained using eddy scattering by surface gravity waves in water. In this case, this is to understand superradiation phenomena from bose-Einstein condensates, \cite{42}. As we've already said, the opposite arrow of this analogy is also interesting. Here, we use this view to reconsider the stability of a constant value vortex in BEC and investigate the special instability that occurs in an inhomogeneous flow BEC similar to a hydrodynamic parallel shear flow.

As we will show clearly now, the Schrodinger-like equation determines the radial function behavior of the space-bounded non-minimum coupled mass scalar field configuration of Kerr-sen black hole space-time, and is suitable for WKB analysis of large mass systems. In particular, Schrodinger-like standard second-order WKB analysis of the radial equation produces the well-known discrete quantization condition(Here we have used the integral relation $\int_{0}^{1} d x \sqrt{1 / x-1}=\pi / 2)$,when $V( r\rightarrow +\infty ),\mu=1/(n+1/2)$, \cite{44}
\begin{equation}
\int_{(y^2)_{t-}}^{(y^2)_{t+}}d(y^2)\sqrt{\omega ^2-V1(y; M^{\prime},a,l,\mu,\alpha1)}=\big(n+{1/2}\big)\cdot\pi\mu /{{2}}=\pi/2
\ \ \ ; \ \ \ \ n=0,1,2,...\  .
\end{equation}
The two integration boundaries $\{y_{t-},y_{t+}\}$ of the WKB
formula are the classical turning points [with
$V(y_{t-})=V(y_{t+})=0$] of the composed
charged-black-hole-massive-field binding potential .
The resonant parameter $n$ (with $n\in\{0,1,2,...\}$) characterizes
the infinitely large discrete resonant spectrum
$\{\alpha_n(\mu,l,M,a)\}_{n=0}^{n=\infty}$ of the black-hole-field
system.

Using the relation  between the radial coordinates $y$ and $r$, one can
express the WKB resonance equation  in the form
\begin{equation}
\int_{r_{t-}}^{r_{t+}}dr{{\sqrt{-V(r; M^{\prime},a,l,\mu,\alpha)}}/{g(r)}}=\big(n+{1/2}\big)\cdot\pi\
\ \ \ ; \ \ \ \ n=0,1,2,...\  ,
\end{equation}
where the two polynomial relations 
\begin{equation}
1-{{2 M^{\prime}}/{r_{t-}}}+{{a^2}/{r^2_{t-}}}=0\
\end{equation}
and
\begin{equation}
{{l(l+1)}/{r^2_{t+}}}+{{2 M^{\prime}}/{r^3_{t+}}}-{{2(a^2)}/{r^4_{t+}}}
-{{\alpha (a^2)}/{r^4_{t+}}}=0\
\end{equation}
determine the radial turning points $\{r_{t-},r_{t+}\}$ of the composed black-hole-field binding potential.

We set
\begin{equation}
x\equiv {{r-r_{\text{+}}}/{r_{\text{+}}}}\ \ \ \ ; \ \ \ \ \tau\equiv {{r_+-r_-}/{r_+}}\  ,
\end{equation}
in terms of which the composed black-hole-massive-field interaction term 
has the form of a binding potential well,
\begin{equation}
V[x(r)]=-\tau\Big({{\alpha(a^2)}/{r^4_+}}-\mu^2\Big)\cdot x +
\Big[{{\alpha(a^2)(5r_+-6r_-)}/{r^5_+}}-\mu^2\big(1-{{2r_-}/{r_+}}\big)\Big]\cdot x^2+O(x^3)\  ,
\end{equation}
in the near-horizon region
\begin{equation}
x\ll\tau\  .
\end{equation}

From the near-horizon expression  of the
black-hole-field binding potential, one obtains the dimensionless
expressions
\begin{equation}
x_{t-}=0\
\end{equation}
and
\begin{equation}
x_{t+}=\tau\cdot{{{{\alpha (a^2)}/{r^4_+}}-\mu^2}/{{{\alpha (a^2)(5r_+-6r_-)}/{r^5_+}}-\mu^2\big(1-{{2r_-}/{r_+}}\big)}}\
\end{equation}
for the classical turning points of the WKB integral relation .

We find that
our analysis is valid in the regime  below($\alpha1$ corresponds to the transformation of y in Equation 38)
\begin{equation}
\alpha\simeq{{\mu^2r^4_+}/{(a^2)}}  ,\alpha1\simeq\sqrt{{{\mu^2r^4_+}/{(a^2)}}}\
\end{equation}
in which case the near-horizon binding potential and its outer
turning point can be approximated by the remarkably compact
expressions
\begin{equation}
V(x)=-\tau\Big[\Big({{\alpha (a^2)}/{r^4_+}}-\mu^2\Big)\cdot
x-4\mu^2\cdot x^2\Big]+O(x^3)\
\end{equation}
and
\begin{equation}
x_{t+}=({1/4})\Big({{\alpha (a^2)}/{\mu^2r^4_+}}-1\Big)\  .
\end{equation}
In addition, one finds the near-horizon relation
\begin{equation}
p(x)=\tau\cdot x+(1-2\tau)\cdot x^2+O(x^3)\  .
\end{equation}

We know that
\begin{equation}
{{1}/{\sqrt{\tau}}}\int_{0}^{x_{t+}}dx \sqrt{{{{\alpha
(a^2)}/{r^2_+}}-\mu^2
r^2_+/{x}}-4\mu^2r^2_+}=\big(n+{1/2})\cdot\pi\ \ \ \ ; \ \ \
\ n=0,1,2,...\  .
\end{equation}
Defining the dimensionless radial coordinate
\begin{equation}
z\equiv {{x}/{x_{t+}}}\  ,
\end{equation}
one can express the WKB resonance equation in the mathematically
compact form
\begin{equation}
({{2\mu r_+ x_{t+}}/{\sqrt{\tau}}})\int_{0}^{1}dz
\sqrt{{{1}/{z}}-1}=\big(n+{1/2})\cdot\pi\ \ \ \ ; \ \ \ \
n=0,1,2,...\  ,
\end{equation}
which yields the relation 
\begin{equation}
{{\mu r_+ x_{t+}}/{\sqrt{\tau}}}=n+{1/2}\ \ \ \ ; \ \ \ \
n=0,1,2,...
\end{equation}

We know from the curve integral formula that there is a certain extreme value forming a loop
\begin{equation}
1/y^2 \rightarrow {\alpha}
\end{equation}y takes the interval from 0 to 1 at this time.

Interestingly, it has been demonstrated numerically in
\cite{16,17} that
the dimensionless physical parameter $\alpha$ {\it diverges} in the
$y$ limit, where the physical parameter
$y$ is defined by the dimensionless relation,for $y$ is greater than $\sqrt{2}$ at this time,
\begin{equation}\label{Eq33}
{y}\equiv{\alpha1}/\sqrt{2} .
\end{equation} 

Here the critical parameter y is given by the
simple relation 
\begin{equation}
{y}/\mu\equiv {{r^2_+}/\sqrt{2(a^2)}}  .
\end{equation}
When
\begin{equation}
\sqrt{2(a^2)}/{r^2_+}< \omega< m\varOmega_H+q\varPhi_H,  
\end{equation}
the Kerr-sen black\ hole\ is\ superradiantly\ stable\ at\ that\ time.

\section{Summary and discussion}
In this article, we introduce $\mu=y\omega$\cite{38} into the Kerr-sen black hole, and discuss the superradiation stability of the Kerr-sen black hole.

In summary, we have studied the stationary massive charged scalar clouds in the Kerr-Sen black hole spacetime. In\cite{39,40}, Hod showed that the Kerr-Newman black hole can support stationary massive charged scalar clouds by analytically solving the Klein-Gordon equation for a stationary charged massive scalar fields. The authors of Ref.\cite{39,40} also numerically investigate stationary massive charged scalar clouds in the Kerr-Newman black hole spacetime and find that for fixed black hole parameters, the mass $\mu$ and charge $q$ of the scalar clouds are limited in a finite region in the parameter space of the scalar fields. The Kerr-Newman scalar clouds were recently promoted to a fully nonlinear solution (Kerr-Newman black holes with scalar hair) in Ref.\cite{39,40}. Of course, the existence of clouds in this Kerr-Sen case shows that a new family of fully nonlinear solutions will also exist in the Kerr-Sen case. This motivated us to perform such an analysis for the Kerr-Sen black hole since Kerr-Newman and Kerr-Sen black holes have several similarities in their physical properties. In this paper, we have studied that the Kerr-Sen black hole can also support stationary massive charged scalar clouds.

In that article\cite{43},when $\mu \ge \sqrt{2}(m\varOmega_H+q\varPhi_H)$,so the Kerr-Newman black hole is superradiantly stable at that time.In this article, a new variable y is added here to expand the results of the above article. When $\sqrt{2(a^2)}/{r^2_+}<\omega< m\varOmega_H+q\varPhi_H,$,so\ the Kerr-sen black\ hole\ is\ superradiantly\ stable\ at\ that\ time.

{\bf Acknowledgements:}\\
I would like to thank Jing-Yi Zhang in GuangZhou University for generous help.This work is partially supported by  National Natural Science Foundation of China(No. 11873025).

\end{document}